\documentclass[publish]{acmconf}
\usepackage{algorithmic}
\usepackage[ps2pdf]{hyperref}
\usepackage{amsmath,amsfonts}
\usepackage[mathscr]{euscript}
\usepackage{dm-math}
\usepackage[latin1]{inputenc}

\CopyrightText{ }

\ConferenceName{28th Annual ACM SIGPLAN-SIGACT Symposium on                             Principles of Programming Languages}
\ConferenceShortName{POPL '01}

\newcommand{\stress}[1]{\textbf{#1}}

\title{An Abstract Monte-Carlo Method for the Analysis of
  Probabilistic Programs\thanks{
This work was partially funded by Commissariat \`a l'\'Energie
Atomique under contract 27234/VSF.}}

\author{\Author{David Monniaux}\\
\Address{\'Ecole Normale Sup\'erieure\\
Laboratoire d'Informatique\\
45, rue d'Ulm\\
75230 Paris cedex 5\\
France\\}
\Email{David.Monniaux@ens.fr}}

\newcommand{\defstress}[1]{\textbf{#1}}

\begin{document}
\maketitle

\begin{abstract}
We introduce a new method, combination of random testing and abstract
interpretation, for the analysis of programs featuring both
probabilistic and non-probabilistic nondeterminism. After introducing
``ordinary'' testing, we show how to combine testing
and abstract interpretation and give formulas linking the precision of the
results to the number of iterations.  We then discuss complexity and
optimization issues and end with some experimental results.
\end{abstract}

\section{Introduction}
We introduce a generic method that lifts an ordinary abstract
interpretation scheme to an analyzer yielding upper bounds on the
probability of certain outcomes, taking into account both randomness
and ordinary nondeterminism.
 
\subsection{Motivations}
It is sometimes desirable to estimate the probability of certain
outcomes of a randomized computation process, such as a randomized
algorithm or an embedded systems whose environment (users, mechanical
and electrical parts\dots) is modeled by known random distributions.
In this latter case, it is particularly important to obtain upper
bounds on the probability of failure.

Let us take an example. A copy machine has a computerized control system that
interacts with the user through some control panel, drives
(servo)motors and receives information from sensors.
In some circumstances, the sensors can give bad information; for
instance, some loose scrap of paper might prevent some optical sensor
from working correctly. It is nevertheless desired that the
probability that the machine will stop in an undesired state (without
having returned the original, for instance) is very low given some
realistic rates of failure from the sensors. To make the system more
reliable, some sensors are redundant and the controlling algorithm
tries to act coherently. Since adding sensors to the design costs
space and hardware, it is interesting to evaluate the probabilities of
failure even before building a prototype.
A similar case can be made of industrial systems such as nuclear power
plants were sensors have a limited life time and cannot be expected to
be reliable. Sound analysis methods are especially needed for that
kind of systems as safety guidelines are often formulated in terms of
maximal probabilities of failures \cite{Leveson86}.

\subsection{Nondeterminism and Probabilities}

Treating the above problem in an entirely probabilistic fashion is not
entirely satisfactory. While it is possible to model the user by
properties such as ``the probability that the user will hit the C key
during the transfer of double-sided documents is less than 1\%'', this
can prevent detecting some failures. For instance, if pressing some
``unlikely'' key combination during a certain phase of copying has a good
chance of preventing correct accounting of the number of copies made, certain
users might use it to get free copies. This is certainly a bug in the
system. To account for the behavior of inputs that cannot be reliably
modeled by random distributions (for instance, malicious attacks)
we must incorporate nondeterminism.
%TODO: blabla sur la différence

\subsection{Comparison to other works}
An important literature has been published on software
\emph{testing} \cite[\dots]{1998:issta:ntafos, Thevenod93};
the purpose of testing techniques is to discover
bugs and even to assert some sort of reliability criterion by testing
the program on a certain number of cases. Such cases are either chosen
randomly \emph{(random testing)} or according to some \emph{ad hoc}
criteria, such as program statement or branch coverage \emph{(partition
testing)}. Partition-based methods can be enhanced by sampling
randomly inside the partition elements.
Often, since the actual distribution in production use is
unknown, a uniform distribution is assumed.

In our case, all the results our method gives are relative to some
fixed, known, distributions driving some inputs. On the other hand, we
will not have to assume some known distribution on the other inputs:
they will be treated as nondeterministic. We thus avoid all problems
pertaining to arbitrary choices of partitions or random distributions;
our method, contrary to most testing methods, is fully mathematically
sound.

There exists a domain called \emph{probabilistic software engineering}
\cite{StatSoftEng}
also aiming at estimating the safety of software. It is based on
statistical studies on syntactic aspects of source code, or software
engineering practices (programming language used, organization of the
development teams\dots), trying to estimate number of bugs in software
according to recorded engineering experience. Our method does not use
such considerations and bases itself on the actual software only.

Our analysis is based on a semantics equivalent to those proposed by
Kozen \cite[2nd semantics]{Kozen79,Kozen81} and Monniaux
\cite{Monniaux_SAS00}.
We proposed a definition of abstract
interpretation on probabilistic programs, using sets of measures, and
gave a generic construction for abstract domains for the construction
of analyzers. Nevertheless, this construction is rather ``algebraic''
and, contrary to the one explained here, does not make use of the
well-studied properties of probabilities.

Ramalingam \cite{PLDI::Ramalingam1996} proposed an abstraction
using vectors of upper bounds of the probabilities of certain
properties, the resulting linear system being solved numerically.
While his approach is sound and effective, it is restricted to
programs where probabilities are only introduced as constant transition
probabilities on the control flow graph. Furthermore, the class of
properties is limited to data-flow analyses.

Several schemes of guarded logic commands \cite{He97} or refinement
\cite{Morgan96} have been introduced. While these systems are
based on semantics broadly equivalent to ours, they are not analysis
systems: they require considerable human input and are rather
formal systems in which to construct derivations of properties of
programs.

\subsection{Contribution}
We introduce for the first time a method combining statistical
and static analyses. This method is proven to be mathematically
sound. While some other methods have been recently proposed to
statically derive properties of probabilistic programs in a general
purpose programming language
\cite{Monniaux_SAS00}, ours is to our knowledge the first that 
makes use of statistical convergences.

\subsection{Structure of the paper}
We shall begin by an explanation of ordinary testing and its
mathematical justification, then explain our ``abstract Monte-Carlo''
method (mathematical bases are given in appendix).
We shall then give the precise concrete semantics that an
abstract interpreter must use to implement our method, first for a
block-structured language then for arbitrary control graphs.
We shall finish with some early results from our implementation.

\section{Abstract Monte-Carlo: the Idea}
In this section, we shall explain, in a mathematical fashion, how our
method works.

\subsection{The Ordinary Monte-Carlo Testing Method}
{\em The reader unfamiliar with probability theory is invited to
consult appendix~\ref{part:measures}.}

Let us consider a deterministic program $c$ whose input $x$ lies in $X$
and whose output lies in $Z$.
We shall note $\semantics{$c$}: X \mapsto Z$ the semantics of $c$
(so that $\semantics{c}(x)$ is the result of the computation of $c$ on
the input $x$). We shall take $X$ and $Z$ two measurable spaces
and constrain $\semantics{$c$}$ to be measurable. These measurability
conditions are technical and do not actually restrict the scope of
programs to consider \cite{Monniaux_SAS00}.
For the sake of simplicity, we shall suppose in this sub-section that
$c$ always terminates.

Let us consider $W \subseteq Z$ a measurable set of
final states whose probability we wish to measure when $x$ is a random
variable  whose probability measure is $\mu$.
The probability of $W$ is therefore $\mu(\inverse{\semantics{$c$}}(W))$.
Noting \[t_W(x) = \begin{cases}
1 & \text{if $\semantics{$c$}(x) \in W$}\\
0 & \text{otherwise,}
\end{cases}\] this probability is the expectation $\expect{t_W}$.

Let us apply the Monte-Carlo method for averages to this random
variable $t_W$ (see appendix~\ref{part:monte-carlo}).
$\expect{t_W}$ is then approximated by $n$ random trials:
\begin{algorithmic}
\STATE $c \leftarrow 0$
\FOR{$i=1$ to $n$}
\STATE $x \leftarrow \textrm{random}(\mu)$
\STATE run program $c$ on input $x$.
\IF{program run ended in a state in $W$}
\STATE $c \leftarrow c+1$
\ENDIF
\ENDFOR
\STATE $p \leftarrow c/n$
\end{algorithmic}
A confidence interval can be supplied, for instance using the
Chernoff-Hoeffding bound (Inequ.~\ref{equ:used}):
there is at least a $1-\varepsilon$
probability that the true expectation $\expect{t_W}$ is less
than $p' = p+\sqrt{\frac{-\log{\varepsilon}}{2n}}$
(Fig.~\ref{fig:prob-exceed} --- we shall see the implications in terms
of complexity of these safety margins in more detail in
section~\ref{part:complexity}).

This method suffers from two drawbacks that make it unsuitable in certain
cases:
\begin{itemize}
\item It supposes that all inputs to the program are either constant
  or driven according to a known probability distribution. In
  general, this is not the case: some inputs might well be only
  specified by intervals of possible values, without any probability
  measure. 
  In such cases, it is common \cite{1998:issta:ntafos}
  to assume some kind of distribution on the inputs, such as an
  uniform one for numeric inputs. This might work in some cases, but
  grossly fail in others, since this is mathematically unsound.

\item It supposes that the program terminates every time within an
  acceptable delay.
\end{itemize}
We propose a method that overcomes both of these problems.

\subsection{Abstract Monte-Carlo}
We shall now consider the case where the inputs of the program are
divided in two: those, in $X$, that follow a random distribution $\mu$
and those that simply lie in some set $Y$. Now $\semantics{$c$}: X
\times Y \rightarrow Z$. The probability we are now trying to quantify
is $\mu \{x \in X \mid \exists y\in Y~ \semantics{$c$} \langle x, y\rangle
\in W\}$. Some technical conditions must be met so that this
probability is well-defined; namely, the spaces $X$ and $Y$ must be
standard Borel spaces \cite[Def.~12.5]{Kechris}.%
\footnote{Let us suppose $X$ and $Y$ are standard Borel
spaces \cite[§12.B]{Kechris}.
$X\times Y$ is thus a Polish space \cite[§3.A]{Kechris} so that the first
projection $\pi_1$ is continuous.
Let $A=\{x \in X \mid \exists y\in Y~ \semantics{$c$} \langle x, y\rangle
\in W\}$; then $A = \pi_1(\inverse{\semantics{$c$}}(W))$.
Since $\semantics{$c$}$ is a measurable function
and $W$ is a measurable set, $\inverse{\semantics{$c$}}(W)$ is a Borel
subset in the Polish space $X \times Y$.
$A$ is therefore analytic \cite[Def.~14.1]{Kechris};
from Lusin's theorem \cite[Th.~21.10]{Kechris}, it is
universally measurable. In particular, it is
$\mu$-measurable \cite[§17.A]{Kechris}. $\mu(A)$ is thus
well-defined.}
Since countable sets, $\bbR$, products of sequences of standard Borel
spaces are standard Borel \cite[§12.B]{Kechris},
this restriction does not concern most semantics.

Noting \[t_W(x) = \begin{cases}
1 & \text{if $\exists y\in Y~ \semantics{$c$}\langle x,y\rangle \in W$}\\
0 & \text{otherwise,}
\end{cases}\]this probability is the expectation $\expect{t_W}$.

While it would be tempting, we cannot use a straightforward
Monte-Carlo method since, in general, $t_W$ is not computable.%
\footnote{Let us take a
Turing machine (or program in a Turing-complete language) $F$.
There exists an algorithmic translation taking $F$ as input and outputting
the Turing machine $\tilde{F}$ computing the total function
$\varphi_{\tilde{F}}$ so that \[\varphi_{\tilde{F}} \langle x,y\rangle =
\begin{cases}
1 & \text{if $F$ terminates in $y$ or less steps on input $x$}\\
0 & \text{otherwise.}
\end{cases}\]
Let us take $X=Y=\bbN$ and $Z=\{0,1\}$ and the program $\tilde{F}$,
and define $t_{\{1\}}$ as before. $t_{\{1\}}(x)=1$ if and only if $F$
terminates on input $x$. It is a classical fact of computability
theory that the $t_{\{1\}}$ function is not computable for all $F$
\cite{ROG}.
}

Abstract interpretation (see appendix~\ref{part:absint}) is a general
scheme for approximated analyses of safety properties of programs.
We use an abstract interpreter to compute a function $T_W: X
\rightarrow \{0,1\}$ testing the following safety property:
\begin{itemize}
\item $T_W(x)=0$ means that no value of $y \in Y$ results in
  $\semantics{$c$}(x,y) \in W$;
\item $T_W(x)=1$ means that some value of $y \in Y$ \emph{may} result in
  $\semantics{$c$}(x,y) \in W$.
\end{itemize}
This means that for any $x$, $t_w(x) \leq T_W(x)$.
Let us use the following algorithm:
\begin{algorithmic}
\STATE $c \leftarrow 0$
\FOR{$i=1$ to $n$}
\STATE $x \leftarrow \textrm{random}(\mu)$
\STATE $c \leftarrow c+T_W(x)$
\ENDFOR
\STATE $p \leftarrow c/n$
\end{algorithmic}
With the same notations as in the previous sub-section:
$t_W^{(c)} \leq T_W^{(n)}$ and thus the confidence interval is still valid:
there is at least a $1-\varepsilon$
probability that the true expectation $\expect{t_W}$ is less
than $p' = p+\sqrt{\frac{-\log{\varepsilon}}{2n}}$.

We shall see in the following section how to build abstract
interpreters with a view to using them for this Monte-Carlo method.

\section{A Concrete Semantics Suitable for Analysis}

From the previous section, it would seem that it is easy to use any
abstract interpreter in a Monte-Carlo method. Alas, we shall now see
that special precautions must be taken in the presence of calls to
random generators inside loops or, more generally, fixpoints.

\subsection{Concrete Semantics}
We have for now spoken of deterministic programs taking one input $x$
chosen according to some random distribution and one input $y$ in some
domain. Calls to random generators (such as the POSIX %TODO: ref
\texttt{drand48()} function) are usually modeled by a sequence of
independent random variables. If a bounded number of calls ($\leq N$)
to such generators is used in the program, we can consider them as
input values: $x$ is then a tuple $\langle x_1,\ldots,x_N, v\rangle$
where $x_1$, \dots, $x_n$ are the values for the generator and $v$ is
the input of the program. If an unbounded number of calls can be made,
it is tempting to consider as an input a countable sequence of values
$(x_n)_{n\in\bbN}$ where $x_1$ is the result of the first call to the
generator, $x_2$ the result of the second call\dots; a formal
description of such a semantics has been made by Kozen
\cite{Kozen79,Kozen81}.

Such a semantics is not very suitable for program
analysis. Intuitively, analyzing such a semantics implies tracking the
number of calls made to number generators. The problem is that such
simple constructs as:
\begin{center}
\tt if (...) \{ random(); \} else \{\}
\end{center}
are difficult to handle: the countings are not synchronized in both
branches.

We shall now propose another semantics, identifying occurrences of
random generators by their program location and loop indices.
The Backus-Naur form of the programming language we shall consider is:
\begin{tabbing}
\textit{instruction} \= ::= \= \textit{elementary}\\
\> $|$ \> \textit{instruction} \texttt{;} \textit{instruction}\\
\> $|$ \> \texttt{if} \textit{boolean\_expr}\\
\>\> \texttt{then} \textit{instruction}\\
\>\>   \texttt{else} \textit{instruction}\\
\>\>   \texttt{endif}\\
\> $|$ \>\texttt{while} \textit{boolean\_expr}\\
\>\>   \texttt{do} \textit{instruction}\\
\>\>   \texttt{done}\\
\end{tabbing}
We leave the subcomponents largely unspecified, as they are not
relevant to our method.
\textit{elementary} instructions are deterministic, terminating basic
program blocks like assignments and simple expression evaluations.
\textit{boolean\_expr} boolean expressions, such as comparisons,
have semantics as sets of acceptable environments. For instance, a
\textit{boolean\_expr} expression can be \verb@x < y + 4@; its
semantics is the set of execution environments where variables
\texttt{x} and \texttt{y} verify the above comparison.
If we restrict ourselves to a finite number $n$ of integer variables,
an environment is just a $n$-tuple of integers.

The denotational semantics of a code fragment $c$ is a mapping from
the set $X$ of possible execution environments before the instruction
into the set $Y$ of possible environments after the instruction. 
Let us take an example. If we take environments as elements of
$\bbZ^3$, representing the values of
three integer variables \texttt{x}, \texttt{y} and \texttt{z}, then
$\semantics{x:=y+z}$ is the strict function
$\langle x, y, z\rangle \mapsto \langle y+z, y, z\rangle$.
Semantics of basic constructs (assignments, arithmetic operators) can
be easily dealt with this forward semantics;
we shall now see how to deal with flow control.

The semantics of a sequence is expressed by simple composition
\begin{equation}
 \semantics{$e_1$; $e_2$} = \semantics{$e_2$} \circ \semantics{$e_1$}
\end{equation}

Tests get expressed easily, using as the semantics $\semantics{$c$}$
of a boolean expression $c$ the set of environments it matches:
\begin{multline}
\semantics{if $c$ then $e_1$ else $e_2$}(x) =\\
  \tifthenelse{x \in \semantics{$c$}}{
    \semantics{$e_1$}(x)}{\semantics{$e_2$}(x)}
\end{multline}
and loops get the usual least-fixpoint semantics (considering the
point-wise extension of the Scott flat ordering on partial functions)
\begin{multline}
  \semantics{while $c$ do $f$} =
  \textrm{lfp}(\lam{\phi}\lam{x} \\ \tifthenelse{x \in \semantics{$c$}}{
     \phi \circ \semantics{$f$}(x)}{x}).
\end{multline}
Non-termination shall be noted by $\bot$.

As for expressions, the only constructs whose semantics we shall
precise are the random generators.
We shall consider a finite set $G$ of different generators. Each generator
$g$ outputs a random variable $r_g$ with distribution $\mu_g$; each
call is independent from the precedent calls.
Let us also consider the set $P$ of program points and the set
$\bbN^*$ of finite sequences of positive integers.
The set $C = P \times \bbN^*$ shall denote the possible times in an
execution where a call to a random generator is made: $\langle p,
n_1n_2...n_l\rangle$ notes the execution of program point $p$ at the
$n_1$-th execution of the outermost program loop, \dots,
$n_l$-th execution of the innermost loop at that point.
$C$ is countable. We shall suppose that inside the inputs of the
program there is for each generator $g$ in $G$ a family
$(\hat{g}_{\langle p,w\rangle})_{\langle p,w\rangle \in C}$ of random
choices.

The semantics of the language then become:
\begin{equation}
\semantics{$e_1$; $e_2$} = \semantics{$e_2$} \circ \semantics{$e_1$}
\end{equation}

Tests get expressed easily, using as the semantics $\semantics{$c$}$
of a boolean expression $c$ the set of environments it matches:
\begin{multline}
  \semantics{if $c$ then $e_1$ else $e_2$}.\langle w, x\rangle =\\
  \tifthenelse{x \in \semantics{$c$}}{
    \semantics{$e_1$}.\langle w, x\rangle}{
    \semantics{$e_2$}.\langle w, x\rangle}
\end{multline}

Loops get the usual least-fixpoint semantics (considering the
point-wise extension of the Scott flat ordering on partial functions):
\begin{multline}
  \semantics{while $c$ do $f$}.\langle w_0, x_0\rangle =\\
  \lfp{\left(\lam{\phi}\lam{\langle w,x\rangle}
     \tifthenelse{x \in \semantics{$c$}}{
     \phi \circ S \circ \semantics{$f$} \langle w,x \rangle)}{x}\right)}.
  \langle 1.w_0,x_0 \rangle
\end{multline}
where $S.\langle c.w,x\rangle = \langle (c+1).w,x\rangle$.
The only change is that we keep track of the iterations of the loop.

As for random expressions,
\begin{equation}
\semantics{$p: \texttt{random}_g$}.\langle w,x\rangle =
\hat{g}_{\langle p, w\rangle}
\end{equation} where $p$ is the program point.

This semantics is equivalent to the denotational
semantics proposed by Kozen \cite[2nd
semantics]{Kozen79,Kozen81} and Monniaux \cite{Monniaux_SAS00},
the semantic of a
program being a continuous linear operator mapping an input measure to
the corresponding output. The key point of this equivalence is that
two invocations of random generators in the same execution have
different indices, which implies that a fresh output of a random
generator is randomly independent of the environment coming to that
program point.

\subsection{Analysis}
Our analysis algorithm is a randomized version of an ordinary
abstract interpreter. Informally, we treat calls to random generators
are treated as follows:
\begin{itemize}
\item calls occurring outside fixpoint convergence iterations are
  interpreted as constants chosen randomly by the interpreter;
\item calls occurring inside fixpoint convergence iterations are
  interpreted as upper approximations of the whole domain of values
  the random generator yield.
\end{itemize}
For instance, in the following C program:
\begin{verbatim}
int x;
x = coin_flip(); /* coin_flip() returns 0 or 1 */
               /* each with probability 0.5 */
for(i=0; i<5; i++)
{
  x = x + coin_flip();
}
\end{verbatim}
the first occurrence of \verb+coin_flip()+ will be treated as a random
value, while the second occurrence will be treated as the least upper
bound of $\{0\}$ and $\{1\}$.

This holds for ``naive'' abstract interpreters; more advanced ones
might perform ``dynamic loop unrolling'' or other semantic
transformations corresponding to a refinement of the abstract domain
to handle execution traces:
{\small
\begin{multline}
\semantics{while $c$ do $e$}(x) =\\
\biggl(
  \biggl( \bigcup_{k < N_1+N_2} \psi^k(x) \biggr) \cup
  \psi^{N_2} \bigl(\lfp \bigl(\lam{l} \psi^{N_1} (x) \cup \psi(l)\bigr)\bigr)
\biggr) \cap \compl{\semantics{$c$}}
\end{multline}%
}%
where $\psi(x) = \semantics{$e$} (x \cap \semantics{$c$})$ and $N_1$
and $N_2$ are possibly decided at run-time, depending on the computed
values.
In this case, the interpreter uses a random generator for the
occurrences of $\texttt{random}_g$ operations outside $\lfp$
computations and abstract values for the operations inside $\lfp$'s.
Its execution defines the finite set $K$ of $\langle p, n_1\dots n_l\rangle$
tags uniquely identifying the random values chosen for
$\hat{g}_{\langle p, n_1\dots n_l\rangle}$, as well as the values
$(\check{g}_c)_{c \in K}$ that have been chosen.
This yields
\begin{multline}
\forall (\hat{g}_c)_{g\in G, c \in C}~
\forall y \in Y~
(\forall c \in K~\hat{g}_c = \check{g}_c) \Rightarrow\\
\semantics{$c$} \langle (\hat{g}_c)_{g\in G, c \in C}, y \rangle \in
\gamma_Z (\abstr{z})
\end{multline}
which means that
\begin{multline}
\forall (\hat{g}_c)_{g\in G, c \in C}~
(\forall c \in K~ \hat{g}_c = \check{g}_c) \Rightarrow\\
t_W((\hat{g}_c)_{g\in G, c \in C}) \leq \tau_W(\abstr{z})
\end{multline}
If we virtually choose randomly some $\check{g}_c$ for $c \notin K$, we
know that $t_W((\check{g}_c)_{g\in G, c \in C}) \leq \tau_W(\abstr{z})$.
Furthermore, $(\check{g}_c)$ follows the product random
distribution $\mu_g^{\otimes C}$ (each $\check{g}_c$ has been chosen
independently of the others according to measure $\mu_g$).

Let us summarize: we wish to generate upper bounds of
experimental averages of a
Bernoulli random variable $t_W: X \rightarrow \{0, 1\}$ whose domain
has the product probability measure $\mu_I \otimes \bigotimes_{g \in G}
\mu_g^{\otimes C}$ where $\mu_I$ is the input measure and the $\mu_g$'s are the
measures for the random number generators. The problem is that the
domain of this random variable is made of countable sequences; thus we
cannot generate its input strictly speaking. We instead effectively choose at
random a finite number of coordinates for the countable sequences, and
compute a common upper bound for $t_W$ for all inputs identical to our
chosen values on this finite number of coordinates. This is identical
to virtually choosing a random countable sequence $x$ and getting an upper
bound of its image by $t_W$.

Implementing such an analysis inside an ordinary abstract interpreter
is easy. The calls to random generators are interpreted as either a
random generation, or as the least upper bound over the range of
the generator, depending on a ``randomize'' flag. This flag is
adjusted depending on whether the interpreter is computing a
fixpoint. The interpreter does not track the indices of the random
variables: these are only needed for the proof of correctness.
The analyzer does a certain number $n$ of trials and outputs the
experimental average $\bar{t}^{(n)}_W$. As a convenience, our
implementation also outputs the $\bar{t}^{(n)}_W+t$ upper bound so
that there is at least a probability $1-\varepsilon$ that this upper
bound is safe according to inequation~(\ref{equ:used}).
This is the value that is reported in the experiments of
section~\ref{sec:experiments}.

While our explanations referred to a forward semantics, the abstract
interpreter can of course combine forward and backward analysis
\cite[section~6]{Cousot92}, provided the chosen random values
are recorded so that subsequent passes of analysis can reuse
them. Another related improvement, explained in section~\ref{part:complexity},
uses a preliminary backward analysis prior to random generation.

\subsection{Arbitrary control-flow graphs}
The abstract interpretation framework can be extended to logic
languages, functional languages and imperative languages with
recursion and other
``complex programming mechanisms
(call-by-reference, local procedures passed as parameters, non-local
gotos, exceptions)'' \cite{BourdonclePhD}.
In such cases, the semantics of the program are expressed as a
fixpoint of a system of equations over parts of the domain of
environments. The environment in that case includes the program
counter, call stack and
memory heap; of course a suitable abstract lattice must be used.

Analyzing a program $P$ written in a realistic imperative language is very
similar to analyzing the following interpreter:
\begin{algorithmic}
\STATE $s \leftarrow \textrm{initial state for $P$}$
\WHILE{$s$ is not a termination state}
\STATE $s \leftarrow N(s)$
\ENDWHILE
\end{algorithmic}
where $N(s)$ is the next-state function for $P$ (operational
semantics). The abstract analyzer analysis that loop using an
abstract state and an abstract version of $N$. Most analyses partition
the domain of states according to the program counter, and the
abstract interpreter then computes the least fixpoint of a system of
semantic equations.

Such an analysis can be randomized in exactly the same fashion as the
one for block-structured programs presented in the previous
section. It is all the same essential to store the generated values in
a table so that backwards analysis can be used.

\section{Complexity}\label{part:complexity}
The complexity of our method is the product of two independent factors:
\begin{itemize}
\item the complexity of one ordinary static analysis of the
  program; strictly speaking, this complexity depends not only on the
  program but on the random choices made, but we can take a rough
  ``average'' estimate that depends only on the program being
  analyzed;

\item the number of iterations, that depends only on the requested
  confidence interval; the minimal number of iterations to reach a certain
  confidence criterion can be derived from inequalities
  \cite[appendix~A]{ShorackWellner} such as inequation
  (\ref{equ:used}) and does not depend on the actual program
  being analyzed.
\end{itemize}
We shall now focus on the latter factor, as the former depends on the
particular case of analysis being implemented.

\begin{figure}
{\fontsize{6}{8}\selectfont
\input{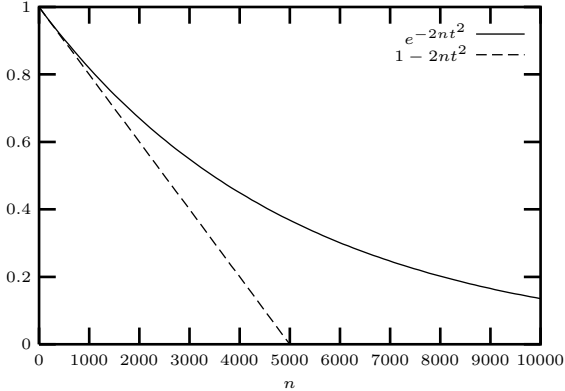}
}
\caption{\label{fig:prob-exceed}
Upper bound on the probability that the computed probability
exceeds the real value by more than $t$, for $t=0.01$.}
\end{figure}

\begin{figure}
{\fontsize{6}{8}\selectfont
\input{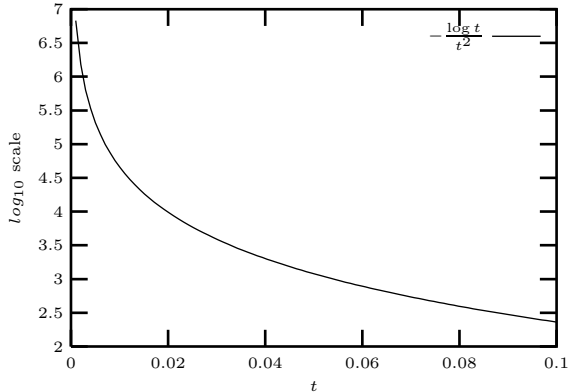}
}
\caption{\label{fig:speed}
Numbers of iterations necessary to achieve a probability of
false report on the same order of magnitude as the error margin.}
\end{figure}

Let us recall inequation~(\ref{equ:used}):
$\pr{\expect{t_W} \geq \bar{t}_W^{(n)} + t} \leq e^{-2nt^2}$. {\em It means that
to get with $1-\varepsilon$ probability an approximation of the requested
probability $\mu$, it is sufficient to compute an experimental average
over $\left\lceil -\frac{\log \varepsilon}{2t^2}\right\rceil$ trials.}

This exponential improvement in quality (Fig.~\ref{fig:prob-exceed})
is nevertheless not that
interesting. Indeed, in practice, we might want $\varepsilon$ and $t$ of
the same order of magnitude as $\mu$.
Let us take $\varepsilon = \alpha t$ where $\alpha$ is fixed.
We then have $n \sim -\frac{\log t}{t^2}$, which indicates prohibitive
computation times for low probability events (Fig.~\ref{fig:speed}).
{\em This high cost of computation for low-probability events is not
specific to our method; it is true of any Monte-Carlo method},
since it is inherent in the speed of convergence of averages of
identically distributed random variables;
this relates to the speed of convergence in the central limit
theorem \cite[ch~1]{ShorackWellner}.
It can nevertheless be circumvented by tricks aimed at estimating the
desired low probability by computing some other, bigger, probability
from which the desired result can be computed.

Fortunately, such an improvement is possible in our method.
If we know that $\pi_1(\semantics{$c$}^{-1}(W)) \subseteq R$, with a
measurable $R$, then we can replace the random variable $t_W$ by its
restriction to $R$: $\restrict{{t_W}}{R}$; then
$\expect t_W = \pr{R}.\expect{\restrict{{t_W}}{R}}$.
If $\pr{R}$ and $\expect t_W$ are on the same order of magnitude, this
means that $\expect{\restrict{{t_W}}{R}}$ will be large and thus that
the number of required iterations will be low.
{\em Such a restricting $R$ can be obtained by static analysis, using
ordinary backwards abstract interpretation.}

A salient point of our method is that our Monte-Carlo computations are
\textbf{highly parallelizable},
with linear speed-ups: $n$ iterations on $1$ machine
can be replaced by $n/m$ iterations on $m$ machines, with very little
communication. Our method thus seems especially adapted for clusters
of low-cost PC with off-the-shelf communication hardware, or even more
distributed forms of computing. Another improvement can be to compute
bounds for several $W$ sets simultaneously, doing common computations
only once.

\section{Practical Implementation and Experiments}
\label{sec:experiments}
We have a prototype implementation of our method, implemented on top
of an ordinary abstract interpreter doing forward analysis using
integer and real intervals. Figures~\ref{fig:exemple_beg}
to~\ref{fig:exemple_end} show various examples for which the
probability could be computed exactly by symbolic integration.
Figure~\ref{fig:exemple_complex} shows a simple program whose 
probability of outcome is difficult to figure out by hand.
Of course, more complex programs can be handled, but the current lack
of support of user-defined functions and mixed use of reals and
integers prevents us from supplying real-life examples. We hope to
overcome these limitations soon as implementation progresses.

\begin{figure}
% exemple2.c
\begin{verbatim}
int x, i;
know (x>=0 && x<=2);
i=0;
while (i < 5)
{
  x += coin_flip();
  i++;
}
know (x<3);
\end{verbatim}

\caption{\textbf{Discrete probabilities.}
The analyzer establishes that, with \textbf{99\% safety}, the
probability $p$ of the outcome ($x < 3$)
is less than \textbf{0.509} given worst-case
nondeterministic choices of the precondition ($x \geq 0 \wedge x \leq
2$). The analyzer used $n=10000$ random trials.
Formally, $p$ is {\small $\pr{\texttt{coin\_flip} \in \{0,1\}^5 \mid
\exists x \in [0,2] \cap \bbZ~\allowbreak
\semantics{$P$}(\texttt{coin\_flip}, x)\allowbreak < 3}$}.
Each \texttt{coin\_flip}
is chosen randomly in $\{0,1\}$ with a uniform distribution.
The exact value is \textbf{0.5}.
}
\label{fig:exemple_beg}
\end{figure}

\begin{figure}
% exemple9.c
\begin{verbatim}
double x;
know (x>=0. && x<=1.);
x+=uniform()+uniform()+uniform();
know (x<2.);
\end{verbatim}

\caption{\textbf{Continuous probabilities.}
The analyzer establishes that, with \textbf{99\%} safety, the
probability $p$ of the outcome ($x < 2$)
is less than \textbf{0.848} given worst-case
nondeterministic choices of the precondition ($x \geq 0 \wedge x \leq
1$). The analyzer used $n=10000$ random trials.
Formally, $p$ is $\pr{\texttt{uniform} \in [0, 1]^3 \mid
\exists x \in [0, 1]~ \semantics{$P$}(\texttt{uniform}, x) < 2}$.
Each \texttt{uniform} is chosen randomly in $[0,1]$ with the
Lebesgue uniform distribution.}
The exact value is $5/6 \approx \mathbf{0.833}$.
\end{figure}

\begin{figure}
% exemple10.c
\begin{verbatim}
double x, i;
know(x<0.0 && x>0.0-1.0);
i=0.;

while (i < 3.0)
{
  x += uniform();
  i += 1.0;
}
know (x<1.0);
\end{verbatim}

\caption{\textbf{Loops.}
The analyzer establishes that, with \textbf{99\% safety}, the
probability $p$ of the outcome ($x < 1$)
is less than $0.859$ given worst-case
nondeterministic choices of the precondition ($x < 0 \wedge x >
-1$). The analyzer used $n=10000$ random trials.
Formally, $p$ is $\pr{\texttt{uniform} \in [0, 1]^3 \mid
\exists x \in [0, 1]~ \semantics{$P$}(\texttt{uniform}, x) < 1}$.
Each \texttt{uniform} is chosen randomly in $[0,1]$ with the
Lebesgue uniform distribution.
The exact value is $5/6 \approx \mathbf{0.833}$.
}
\label{fig:exemple_end}
\end{figure}

\begin{figure}
\begin{verbatim}
{
  double x, y, z;
  know (x>=0. && x<=0.1);
  z=uniform(); z+=z;
  if (x+z<2.)
  {
    x += uniform();
  } else
  {
    x -= uniform();
  }
  know (x>0.9 && x<1.1);
}
\end{verbatim}
\caption{%
The analyzer establishes that, with \textbf{99\% safety}, the
probability $p$ of the outcome ($x > 0.9 \wedge x < 1.1$)
is less than \textbf{0.225} given worst-case
nondeterministic choices of the precondition ($x \geq 0 \wedge x
\leq 0.1$). Formally, $p$ is $\pr{\texttt{uniform} \in [0, 1]^2 \mid
\exists x \in [0, 0.1]~ \semantics{$P$}(\texttt{uniform}, x) \in [0.9,1.1]}$.
Each \texttt{uniform} is chosen randomly in $[0,1]$ with the
Lebesgue uniform distribution.
}
\label{fig:exemple_complex}
\end{figure}

\section{Conclusions}
We have proposed a generic method that combines the well-known
techniques of abstract interpretation and Monte-Carlo program testing
into an analysis scheme for probabilistic and nondeterministic
programs, including reactive programs whose inputs are modeled by
both random and nondeterministic inputs. This method is mathematically
proven correct, and uses no assumption apart from the distributions and
nondeterminism domains supplied by the user. It yields upper bounds on
the probability of outcomes of the program, according to the supplied
random distributions, with worse-case behavior according to the
nondeterminism; whether or not this bounds are sound is probabilistic,
and a lower-bound of the soundness of those bounds is supplied. 
While our explanations are given using a simple imperative language as an
example, the method is by no means restricted to imperative programming.

The number of trials, and thus the complexity of the computation,
depends on the desired precision. The method is parallelizable with
linear speed-ups. The complexity of the analysis, or at least its part
dealing with probabilities, increases if the probability to be
evaluated is low. However, static analysis can come to help to reduce
this complexity.

We have implemented the method on top of a simple static analyzer and
conducted experiments showing interesting results on small programs
written in an imperative language. As implementation progresses, we
expect to have results on complex programs akin to those used in
embedded systems.

%\bibliographystyle{plain}
%\bibliography{set-theory,semantics,absint,probas,calcul,testing,David_Monniaux}

\begin{thebibliography}{10}

\bibitem{BourdonclePhD}
François Bourdoncle.
\newblock {\em Sémantiques des Langages Impératifs d'Ordre Supérieur et
  Interprétation Abstraite}.
\newblock PhD thesis, École Polytechnique, 1992.

\bibitem{Cousot92}
Patrick Cousot and Radhia Cousot.
\newblock Abstract interpretation and application to logic programs.
\newblock {\em J. Logic Prog.}, 2-3(13):103--179, 1992.

\bibitem{CousotHalbwachs78}
Patrick Cousot and Nicolas Halbwachs.
\newblock Automatic discovery of linear restraints among variables of a
  program.
\newblock In {\em Proceedings of the Fifth Conference on Principles of
  Programming Languages}. ACM Press, 1978.

\bibitem{Doob}
J.L. Doob.
\newblock {\em Measure Theory}, volume 143 of {\em Graduate Texts in
  Mathematics}.
\newblock Springer-Verlag, 1994.

\bibitem{He97}
Jifeng He, K.~Seidel, and A.~McIver.
\newblock Probabilistic models for the guarded command language.
\newblock {\em Science of Computer Programming}, 28(2--3):171--192, April 1997.
\newblock Formal specifications: foundations, methods, tools and applications
  (Konstancin, 1995).

\bibitem{Hoeffding63}
Wassily Hoeffding.
\newblock Probability inequalities for sums of bounded random variables.
\newblock {\em J. Amer. Statist. Assoc.}, 58(301):13--30, 1963.

\bibitem{Kechris}
Alexander~S. Kechris.
\newblock {\em Classical descriptive set theory}.
\newblock Graduate Texts in Mathematics. Springer-Verlag, New York, 1995.

\bibitem{Kozen79}
D.~Kozen.
\newblock Semantics of probabilistic programs.
\newblock In {\em 20th Annual Symposium on Foundations of Computer Science},
  pages 101--114, Long Beach, Ca., USA, October 1979. IEEE Computer Society
  Press.

\bibitem{Kozen81}
D.~Kozen.
\newblock Semantics of probabilistic programs.
\newblock {\em Journal of Computer and System Sciences}, 22(3):328--350, 1981.

\bibitem{Leveson86}
N.~G. Leveson.
\newblock Software safety: Why, what, and how.
\newblock {\em Computing Surveys}, 18(2):125--163, June 1986.

\bibitem{Monniaux_SAS00}
David Monniaux.
\newblock Abstract interpretation of probabilistic semantics.
\newblock In {\em Seventh International Static Analysis Symposium (SAS'00)},
  number 1824 in Lecture Notes in Computer Science. Springer-Verlag, 2000.
\newblock © Springer-Verlag.

\bibitem{Morgan96}
Carroll Morgan, Annabelle McIver, Karen Seidel, and J.~W. Sanders.
\newblock Refinement-oriented probability for {CSP}.
\newblock {\em Formal Aspects of Computing}, 8(6):617--647, 1996.

\bibitem{1998:issta:ntafos}
Simeon Ntafos.
\newblock On random and partition testing.
\newblock In Michal Young, editor, {\em {ISSTA}~98: Proceedings of the {ACM}
  {SIGSOFT} International Symposium on Software Testing and Analysis}, pages
  42--48, 1998.

\bibitem{StatSoftEng}
Panel on~Statistical Methods~in Software~Engineering.
\newblock {\em Statistical Software Engineering}.
\newblock National Academy of Sciences, 1996.

\bibitem{PLDI::Ramalingam1996}
G.~Ramalingam.
\newblock Data flow frequency analysis.
\newblock In {\em Proceedings of the {ACM} {SIGPLAN}~'96 Conference on
  Programming Language Design and Implementation}, pages 267--277,
  Philadelphia, Pennsylvania, 21--24~May 1996.

\bibitem{ROG}
H.~Rogers.
\newblock {\em Theory of recursive and effective computability}.
\newblock MGH, 1967.

\bibitem{Rudin}
Walter Rudin.
\newblock {\em Real and Complex Analysis}.
\newblock McGraw-Hill, 1966.

\bibitem{ShorackWellner}
Galen~R. Shorack and Jon~A. Wellner.
\newblock {\em Empirical Processes with Applications to Statistics}.
\newblock Wiley series in probability and mathematical statistics. John Wiley
  \& Sons, 1986.

\bibitem{Thevenod93}
P.~Thévenod-Fosse and H.~Waeselynck.
\newblock Statemate applied to statistical software testing pages 99-109.
\newblock In {\em Proceedings of the 1993 international symposium on Software
  testing and analysis}, pages 99--109. Association for Computer Machinery,
  June 1993.

\end{thebibliography}

\appendix

\section{Probability theory}\label{part:measures}

Throughout this paper we take the usual mathematical point of view of
considering probabilities to be given by \defstress{measures} over
\defstress{measurable sets} \cite{Rudin,Doob}.

\begin{itemize}
\item A \defstress{$\sigma$-algebra}
is a set of subsets of a set $X$ that contains
$\emptyset$ and is stable by countable union and
complementation (and thus contains $X$ and is stable by countable
intersection). For technical reasons, not all sets can be measured
(that is, given a probability) and we have to restrict ourselves to
some sufficiently large $\sigma$-algebras, such as the Borel or
Lebesgue sets \cite{Rudin}.

\item A set $X$ with a $\sigma$-algebra $\sigma_X$
defined on it is called a
\defstress{measurable space} and the elements of the $\sigma$-algebra
are the \defstress{measurable subsets}. We shall often mention
measurable spaces by their name, omitting the $\sigma$-algebra, if no
confusion is possible.

\item If $X$ and $Y$ are measurable spaces, $f: X \rightarrow Y$ is a
\defstress{measurable function} if for all $W$ measurable in $Y$,
$f^{-1}(W)$ is measurable in $X$.

\item A \defstress{positive measure} is a function $\mu$ defined on a
$\sigma$-algebra  $\sigma_X$ whose range is in $[0,\infty]$ and which
is countably additive.
$\mu$ is countably additive if, taking $(A_n)_{n\in\bbN}$ a
disjoint collection of elements of $\sigma_X$, then
$\mu\left(\cup_{n=0}^\infty A_n\right) = \sum_{n=0}^\infty \mu(A_n)$.
To avoid trivialities, we assume $\mu(A) < \infty$ for at least one
$A$.
%The \defstress{total weight} of a measure $\mu$ is
%$\mu(X)$. $\mu$ is said to be \defstress{concentrated} on $A \subseteq
%X$ if for all $B$, $\mu(B) = \mu(B \cap A)$. We shall note
%$\measures{X}$ the positive measures on $X$.

If $X$ is countable, $\sigma_X$ can be $\parts{X}$, the power-set of
$X$, and a measure $\mu$ is determined by its value on the singletons:
for any $A \subseteq X$, $\mu(A) = \sum_{a \in A} \mu(\{a\})$.

% \item A \defstress{$\sigma$-finite measure} on $X$
% is a measure $\mu$ so that there exists a countable family of
% measurable sets $(A_n)_{n\in\bbN}$ so that $\forall n~\mu(A_n) <
% \infty$ and $\bigcup_n A_n = X$. We note by $\measures{X}$
% the $\sigma$-finite measures on $X$.

\item A \defstress{probability measure} is a positive measure of total
weight $1$; a \defstress{sub-probability measure} has total weight less
or equal to $1$. We shall note $\pmeasures{X}$ the sub-probability
measures on $X$.

\item Given two sub-probability measures $\mu$ and $\mu'$ (or more
generally, two $\sigma$-finite measures) on $X$ and $X'$ respectively,
we note $\mprod{\mu}{\mu'}$
the product measure \cite[definition 7.7]{Rudin}, defined on the 
product $\sigma$-algebra $\sigma_X \times \sigma_{X'}$.
The characterizing
property of this product measure is that $\mu \otimes \mu' (A \times
A') = \mu(A) . \mu'(A')$ for all measurable sets $A$ and $A'$.
It is also possible to define countable products of measures;
if $\mu$ is a measure over the measurable space $X$, then
$\mu^{\otimes \bbN}$ is a measure over the set $X^\bbN$ of sequences
of elements of $X$.
\end{itemize}

For instance,
let us take $\mu$ the measure on the set $\{0,1\}$ with $\mu(\{1\})=p$
and $\mu(\{0\})=1-p$. Let us take $S$ the set of sequences over
$\{0,1\}$ beginning with $\langle 0,0,1,0\rangle$.
$\mu^{\otimes \bbN}(S) = p(1-p)^3$ is the probability of getting
a sequence beginning with $\langle 0,0,1,0\rangle$
when choosing at random a countable sequence of $\{0,1\}$ independently
distributed following~$\mu$.

\section{Estimating the probability of a random event by the
  Monte-Carlo method}
\label{part:monte-carlo}

We consider a system whose outcome (success or failure) depends on the
value of a parameter $x$, chosen in the set $X$
according to a random distribution $\mu$. The behavior of this system
is described by a random variable $V: X \rightarrow \{0,1\}$, where
$0$ means success and $1$ failure.

The \stress{law of large numbers} says that if we independently choose inputs
$x_k$, with distribution $\mu$, and compute the experimental average
$V^{(n)} = \frac{1}{n} \sum_{k=1}^n V(x_k)$, then
$\lim_{n \rightarrow \infty} V^{(n)} = \expect{V}$
% TODO: préciser signification de la convergence (random variables)
where $\expect{V}$ is the expectation of failure.
Intuitively, it is possible to estimate accurately $\expect{V}$ by
effectively computing $V^{(n)}$ for a large enough value of $n$.

\begin{figure}
\begin{center}
\input{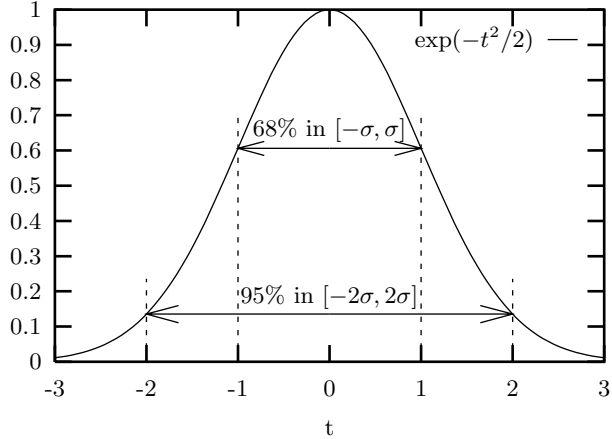}
\end{center}
\caption{The Gaussian normal distribution centered on $0$, with standard
deviate $1$.}
\label{fig:gaussian}
\end{figure}

Just how far should we go? Unfortunately, a general feature of all
Monte-Carlo methods is their slow asymptotic convergence speed.
Indeed, the distribution of the experimental average $V^{(n)}$
is a binomial distribution centered around $\expect{V}$. With
large enough values of $n$ (say $n \geq 20$), this binomial distribution
behaves mostly like a normal (Gaussian) distribution
(Fig.~\ref{fig:gaussian}) with means $p=\expect{V}$
and standard deviate $\frac{p(1-p)}{\sqrt{n}}$. More generally, the
central limit theorem predicts that the average of $n$ random
variables identically distributed as $V$ % TODO: hypothesis
%TODO: citer théorème
has the same expectation $\expect{V}$ as $V$ and standard deviate
$\frac{\sigma}{\sqrt{n}}$ where $\sigma$ is the standard deviate of
$V$. The standard deviate measures the error margin on the computed
result: samples from a gaussian variable centered on $x_0$ and with
standard deviate $\sigma$ fall within $[x_0-2\sigma,x_0+2\sigma]$
about 95\% of the time.

We can better evaluate the probability of underestimating the
probability by more than $t$ using the Chernoff-Hoeffding
\cite{Hoeffding63} \cite[inequality~A.4.4]{ShorackWellner} bounds:
\begin{equation}\label{equ:used}
\pr{\expect{V} \geq V^{(n)} + t} \leq e^{-2nt^2}
\end{equation}
This bound, fully mathematically sound, means that the
probability of underestimating $V$ using $V^{(n)}$ by more than $t$ is
less than $e^{-2nt^2}$.

Any Monte-Carlo method has an inherent margin of error; this
margin of error is probabilistic, in the sense that facts such as
``the value we want to compute is in the interval $[a,b]$'' are valid
up to a certain probability. The size of the interval of safety for a
given probability of error varies in $1/\sqrt{n}$.

\section{Abstract Interpretation}\label{part:absint}
Let us recall the mathematical foundations of abstract
interpretation \cite{CousotHalbwachs78,Cousot92}.	
Let us consider two preordered sets $\abstr{A}$ and $\abstr{Z}$ so
that there exist monotone functions $\gamma_A: \abstr{A} \rightarrow
\parts{A}$, where $A=X\times Y$, and
$\gamma_W: \abstr{Z} \rightarrow \parts{Z}$,
where $\parts{Z}$ is the set of parts of set $Z$, ordered by
inclusion. $\gamma_W$ is called the \defstress{concretization
function}.

The elements in $\abstr{A}$ and $\abstr{Z}$ represent some
properties; for instance, if $X=\bbZ^m$ and $Y=\bbZ^n$, $\abstr{A}$
could be the set of machine descriptions of polyhedra in $\bbZ^{m+n}$ and
$\gamma_A$ the function mapping the description to the set of points
inside the polyhedron \cite{CousotHalbwachs78}. A simpler case is the
intervals, where the machine description is an array of integer couples
$\langle a_1, b_1; a_2, b_2; \dots ; a_n, b_n \rangle$ and its
concretization is the set of tuples $\langle c_1; \dots; c_n \rangle$
where for all $i$, $a_i \leq c_i \leq b_i$.

We then define an \defstress{abstract interpretation} of program $c$
to be a monotone function $\abstr{\semantics {$c$}}: \abstr{A} \rightarrow
\abstr{Z}$ so that
\[\forall \abstr{a} \in \abstr{A},~ \forall a \in A~
a \in \gamma_A(\abstr{A}) \Rightarrow
\semantics{$c$}(a) \in \gamma_Z \circ \abstr{\semantics{$c$}}(\abstr{a}).\]

In the case of intervals, abstract interpretation propagates intervals
of variation along the computation. Loops get a fixpoint semantics:
the abstract interpreter heuristically tries to find intervals that are
invariant for the body of the loop. Such heuristics are based on
widening and narrowing operators \cite{Cousot92}.

It is all the same possible to define backwards abstract
interpretation: a \defstress{backwards abstract interpretation}
of a program $c$
is a monotone function $\abstr{{\semantics {$c$}^{-1}}}: \abstr{Z} \rightarrow
\abstr{A}$ so that
\[\forall \abstr{z} \in \abstr{Z},~ \forall z \in Z~
z \in \gamma_Z(\abstr{Z}) \Rightarrow
\semantics{$c$}^{-1}(z) \subseteq
\gamma_A \circ \abstr{{\semantics{$c$}^{-1}}}(\abstr{z}).\]
Further refinement can be achieved by iterating forwards and backwards
abstract interpretations \cite{Cousot92}.

\end{document}